\documentstyle[epsf,twoside,fleqn,espcrc2]{article}

\newcommand{\AmS}{{\protect\the\textfont2
  A\kern-.1667em\lower.5ex\hbox{M}\kern-.125emS}}

\def\simm#1{\mathop{\vtop{\ialign{##\crcr
        $\hfil\displaystyle{#1}\hfil$\crcr\noalign{\kern0.5pt\nointerlineskip}
        $\sim$\crcr\noalign{\kern0.5pt}}}}\limits}

\def\be{\begin{equation}}
\def\ee{\end{equation}}
\def\ba{\begin{eqnarray}}
\def\ea{\end{eqnarray}}

\def\P{{\sf P}}
\def\R{{\sf R}}
\def\W{{\sf W}}
\def\C{{\sf C}}
\def\PW{{\sf P-W}}
\def\PC{{\sf P-C}}
\def\RW{{\sf R-W}}
\def\RC{{\sf R-C}}

\title{
\vspace*{-32pt}
{\normalsize \hfill UTHEP-369} \\
\vspace*{-5pt}
{\normalsize \hfill UTCCP-P-22} \\
\vspace*{-5pt}
{\normalsize \hfill August 1997} \\
Full QCD simulation on CP-PACS
\thanks{Talk presented by K.\ Kanaya at the International Workshop on 
``LATTICE QCD ON PARALLEL COMPUTERS'', 10-15 March 1997, Center for
Computational Physics, University of Tsukuba.}
}

\author{
CP-PACS Collaboration:\\
\vspace{2mm}
S.\ Aoki\rlap,\address{Institute of Physics,
University of Tsukuba, Tsukuba, Ibaraki 305, Japan}
G.\ Boyd\rlap,\address{Center for Computational Physics, 
University of Tsukuba, Tsukuba, Ibaraki 305, Japan}
R.\ Burkhalter\rlap,$^{\rm b}$
S.\ Hashimoto\rlap,\address{Computing Research Center, 
High Energy Accelerator Research Organization (KEK), Tsukuba, Ibaraki 305, 
Japan}
N.\ Ishizuka\rlap,$^{\rm a}$
Y.\ Iwasaki\rlap,$^{\rm a,b}$
K.\ Kanaya\rlap,$^{\rm a,b}$
T.\ Kaneko\rlap,$^{\rm a}$
Y.\ Kuramashi\rlap,\address{Institute of Particle and Nuclear Studies,
High Energy Accelerator Research Organization (KEK), Tsukuba, Ibaraki 305, 
Japan}
M.\ Okawa\rlap,$^{\rm d}$
A.\ Ukawa\rlap,$^{\rm a}$
and
T.\ Yoshi\'e$^{\rm a,b}$
}

\begin{document}
\renewcommand{\textfraction}{0.1}
\renewcommand{\topfraction}{0.9}
\begin{abstract}
A status report is made of an on-going full QCD study on the CP-PACS aiming 
at a comparative analysis of the effects of improving gauge and quark actions on 
hadronic quantities and static quark potential.  Simulations are made 
for four action 
combinations, the plaquette or an RG-improved action for gluons and 
the Wilson or SW-clover action for quarks, at  
$a^{-1} \approx 1.1$-1.3GeV and $m_\pi/m_\rho \approx 0.7$-0.9.  
Results demonstrate clearly that 
the clover term markedly reduces discretization
errors for hadron spectrum, while adding six-link terms to the plaquette 
action leads to much better rotational symmetry in the potential.  These
results extend experience with quenched simulations to full QCD.
\end{abstract}

\maketitle
\setcounter{footnote}{0}

\section{Introduction}
\label{sect:intro}
\vspace{-1mm}

With the progress in recent years of quenched simulations of QCD, 
deviations of the quenched hadron spectrum from experiment 
are being uncovered.
For heavy quark systems precise calculations with NRQCD have shown 
that the fine structure of quarkonium spectra can be reproduced only if sea 
quark effects are taken into account \cite{NRQCD97}. 
For the light hadron sector several reports have been made 
that strange quark mass cannot be set consistently from 
pseudo scalar and vector meson channels in quenched QCD 
\cite{UKQCDJ,LosAlamos96,UKQCDKphi,YoshieCCP,UKQCDkenway}. 
Most recently results of an extensive quenched 
simulation on the CP-PACS indicate that there is a systematic disagreement 
in the spectrum of baryons \cite{YoshieCCP}.  
Clearly the time has come to bolster efforts toward full QCD simulations. 
We have recently started an attempt in this direction using the CP-PACS 
computer \cite{cppacs}.

Full QCD simulations are, however, extremely computer time consuming 
compared to those of quenched QCD.  
Even with the TFLOPS-class computers that are 
becoming available, high statistics studies, indispensable for reliable 
results, will be difficult for lattice sizes exceeding $32^3\times 64$.
Since a physical lattice size of 
$L\approx 2.5$-3.0fm is needed to avoid finite-size effects
\cite{FukugitaAoki,MILC_fullKS,GF11},  
the smallest lattice spacing one can reasonably reach will be 
$a^{-1}\approx 2$GeV.  Hence lattice discretization errors have to 
be controlled with simulations carried out 
at an inverse lattice spacing smaller than this value.  
This will be a difficult task with the standard plaquette 
and Wilson quark actions 
since discretization errors are of order 20-30\% even at 
$a^{-1}\approx 2$GeV \cite{Sloan97}.  
This leads us to consider improved actions for our simulation of full QCD.

Studies of improved actions have been widely pursued in the last few 
years. 
Detailed tests of improvement for hadron spectrum, 
however, have been mostly carried out within 
quenched QCD (see, {\it e.g.}, Refs.~\cite{cornell,scri,bock,ImQQcdMILC}), 
and only a few studies are available
for the case of full QCD \cite{ImFQcdSCRI}.
In particular a systematic investigation 
of how various terms added to the gauge and quark actions, taken separately,
affect light hadron observables has not been carried out in full QCD.  
We have decided to undertake such a study as the first subject of our full 
QCD program.  In this article we report preliminary results of this 
on-going attempt.

\section{Choice of action and simulation parameters}
\label{sect:parameters}
\vspace{-1mm}

\begin{table} [t]
\setlength{\tabcolsep}{0.3pc}
\caption{Simulation parameters of full QCD runs on a $12^3\times32$
lattice. 
Numbers in parenthesis denote number of configurations stored for 
static quark potential analyses. 
}
\label{tab:parameters}
\vspace{-0mm}
\begin{center}  
\begin{tabular}{clllll} 
\hline
 & $\beta$ & $K$ & $C_{SW}$ & 
\#config & $m_\pi/m_\rho$ \\
\hline
\PW  & 4.8 & .1846 &       & 222 (0)   & .83 \\ 
     &     & .1874 &       & 200 (0)   & .77 \\
     &     & .1891 &       & 200 (0)   & .70 \\
     & 5.0 & .1779 &       & 300 (100) & .85 \\
     &     & .1798 &       & 301 (101) & .79 \\
     &     & .1811 &       & 301 (101) & .71 \\
\hline
\RW  & 1.9 & .1632 &       & 200 (0)   & .90 \\
     &     & .1688 &       & 200 (0)   & .80 \\
     &     & .1713 &       & 200 (0)   & .69 \\
     & 2.0 & .1583 &       & 300 (100) & .90 \\
     &     & .1623 &       & 300 (100) & .83 \\
     &     & .1644 &       & 300 (100) & .74 \\
\hline
\PC  & 5.2 & .139  & 1.69  & 208 (100) & .83 \\
     &     & .141  & 1.655 & 207 (100) & .79 \\
     &     & .142  & 1.64  & 200 (100) & .73 \\
     & 5.25& .139  & 1.637 & 198 (0)   & .83 \\
     &     & .141  & 1.61  & 194 (0)   & .76 \\
\hline
\RC  & 2.0 & .1300 & 1.54  & 201 (191) & .90 \\
     &     & .1370 & 1.52  & 200 (190) & .79 \\
     &     & .1388 & 1.515 & 138 (138) & .71 \\
\hline
\end{tabular}
\end{center}
\vspace{-5mm}
\end{table}

Improving the standard plaquette action for gluons requires the addition of
Wilson loops of six links or more in length.  The precise forms of the added
terms and their coefficients differ depending on the principle one follows
for improvement.  In our study we choose an action given by
\be
S_g^{R} = {\beta \over 6}\left(c_0 \sum W_{1\times 1}
               + c_1 \sum W_{1\times 2}\right),
\label{eq:Raction} 
\ee
with $c0=1-8c_1$ and $c_1=-0.331$, which was obtained 
by a renormalization group treatment \cite{Iwasaki83}. 
The quenched static quark potential calculated with this action exhibits 
good rotational symmetry and scaling already at 
$a^{-1}\approx 1$GeV\cite{IwasakiPot97}, 
similar to those observed for tadpole-improved and fixed 
point actions \cite{TIPot,FPPot}.

For improving the quark action we take the clover improvement due to
Sheikholeslami and Wohlert \cite{clover} defined by
\be
D_{xy}^{C}  =  D_{xy}^{W} 
 + \, \delta_{xy} c_{SW} K \sum_{\mu,\nu} 
         \frac{i}{4} \sigma_{\mu,\nu} F_{\mu,\nu}
\label{eq:Caction}
\ee
with the standard Wilson matrix $D_{xy}^{W}$ given by 
\ba
D_{xy}^{W} & = & \delta_{xy} 
- K \sum_\mu \{(1-\gamma_\mu)U_{x,\mu} \delta_{x+\hat\mu,y} \nonumber \\
      & &  +\, (1+\gamma_\mu)U_{x,\mu}^{\dag} \delta_{x,y+\hat\mu} \},
\label{eq:Waction}
\ea
For the clover coefficient $c_{SW}$, we adopt the meanfield
improved value \cite{meanfield}:
$c_{SW} = P^{-3/4}$ with $P$ the plaquette average.

\begin{figure*}[t]
\vspace{-1mm}
\begin{center}
\leavevmode
\epsfxsize=7.5cm \epsfbox{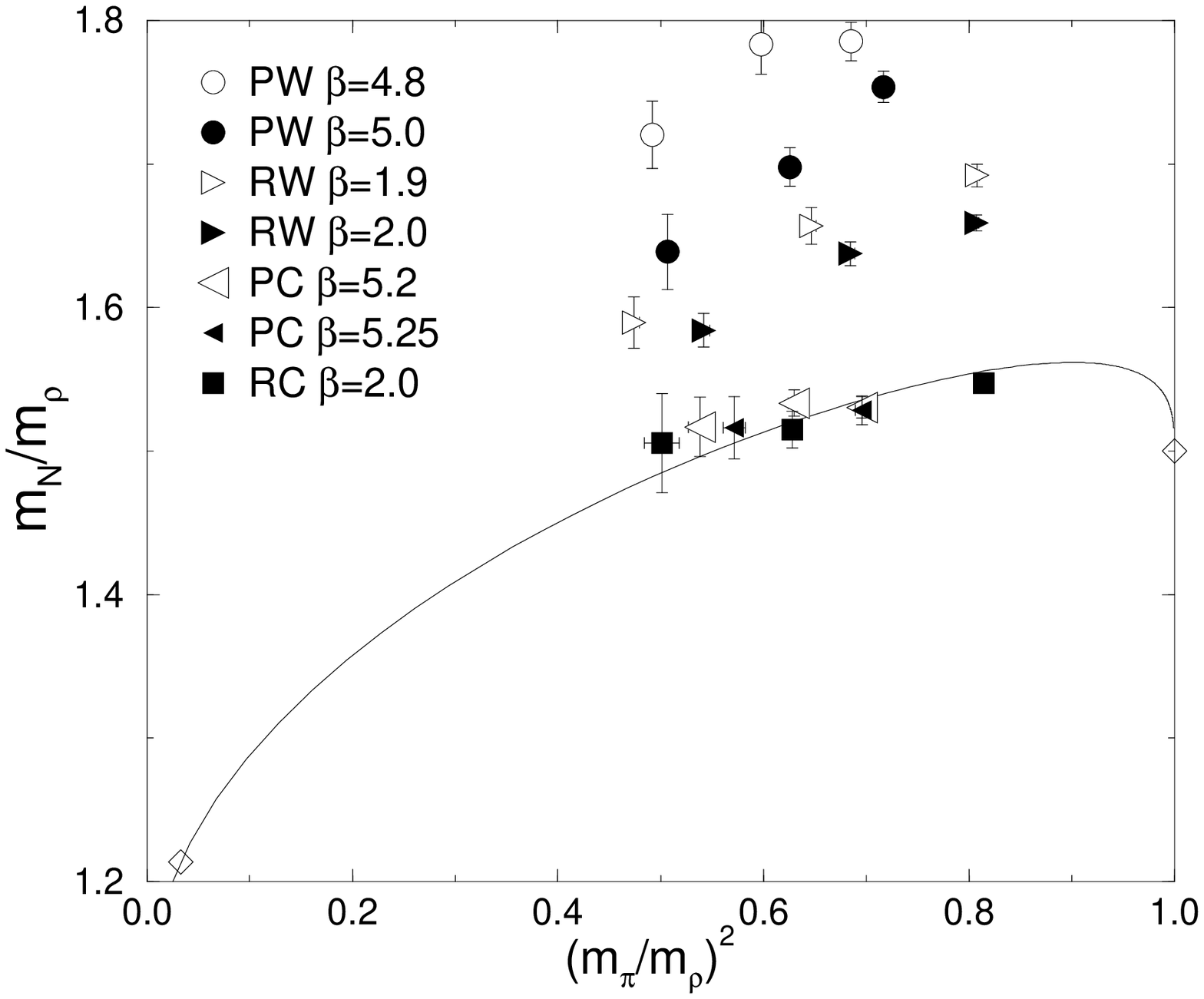}
\hspace{3mm}
\epsfxsize=7.5cm \epsfbox{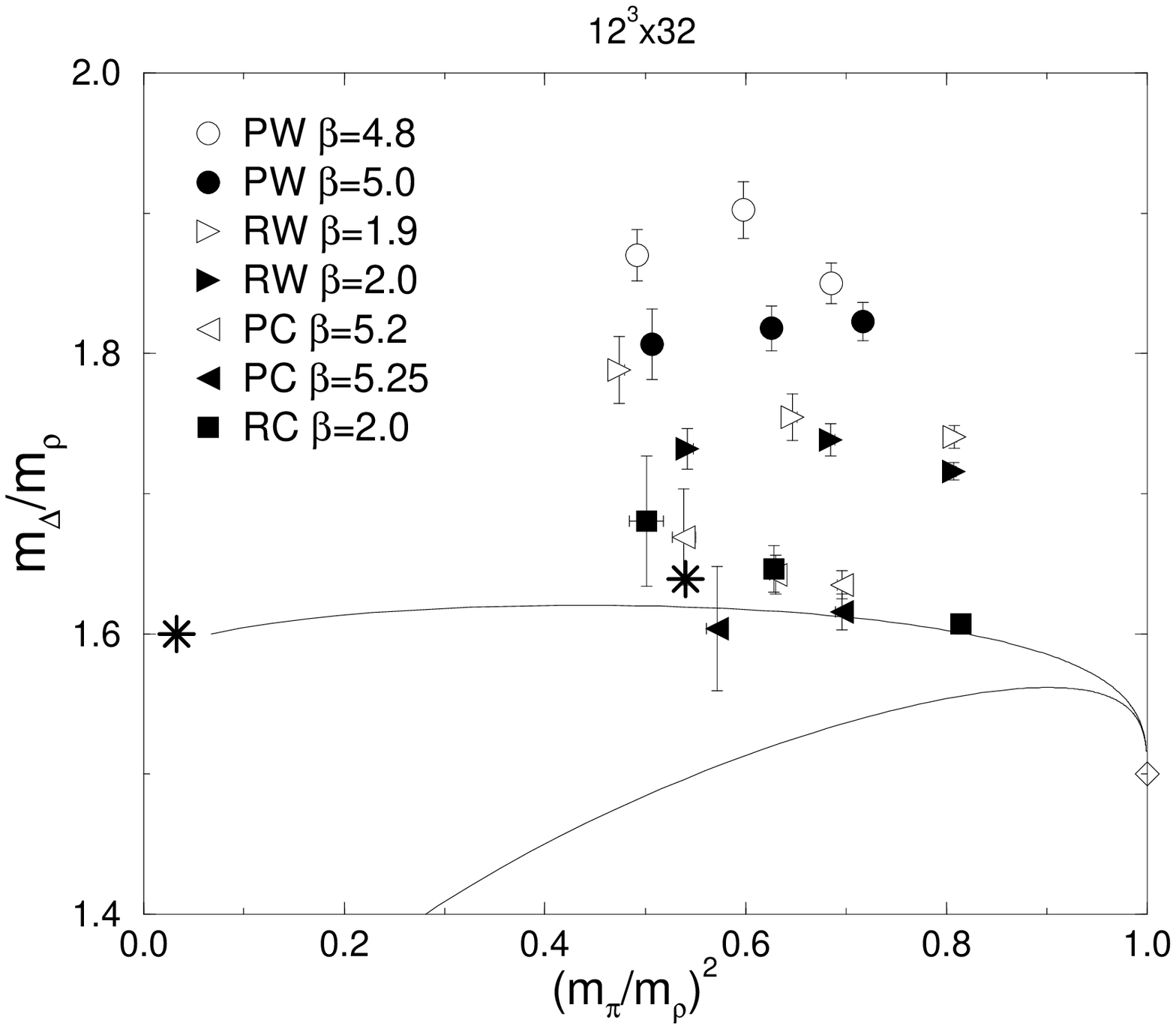}
\end{center}
\vspace{-17mm}
\caption{
(a) $m_N/m_\rho$ and (b) $m_\Delta/m_\rho$ as a function of 
$(m_\pi/m_\rho)^2$ for various combinations of the action.
Stars in (b) are experimental points corresponding to 
$\Delta(1232)/\rho(770)$ and $\Omega(1672)/\phi(1020)$.
}
\label{fig:EPN}
\vspace{-4mm}
\end{figure*}

We carry out a comparative study of the light hadron spectrum and static quark
potential for four action combinations, choosing either the plaquette (\P)
or the rectangular action above (\R) for gluons and either the Wilson (\W)
or clover action (\C) for quarks.
We expect the extent of improvement to 
be clearer at a coarser lattice spacing.  We therefore attempt to 
tune the coupling constant $\beta$ so that the lattice spacing 
equals $a^{-1} \sim 1$GeV.

Our simulations are carried out for two flavors of quarks, 
mostly on a $12^3\times 32$ lattice with 
additional runs on an $8^3\times 16$ lattice to estimate parameters, 
including a self-consistent value of $c_{SW}$.  
We employ the hybrid 
Monte Carlo algorithm to generate full QCD configurations at two or three 
values of $K$ corresponding to $m_\pi/m_\rho \approx 0.7$-0.9. 
The molecular dynamics step size is chosen 
to yield an acceptance of 80-90\%.   
After thermalizing for 100-200 trajectories we generate 1000-1500 
trajectories.

We measure hadron propagators every 5 trajectories, using point and smeared 
sources and point sinks following the method of our quenched 
study \cite{YoshieCCP}. 
The static quark potential is calculated on a subset of 
configurations used for hadron propagator measurement.
The smearing technique of Ref.~\cite{StdST} is employed  
choosing the number of smearing steps and fitting ranges from experience 
in Ref.~\cite{IwasakiPot97}.
Errors are estimated by a single-elimination jackknife procedure.
Simulation parameters of our runs and 
the number of configurations used for the spectrum and potential 
measurements are summarized in Table~\ref{tab:parameters}.

\begin{figure*}[t]
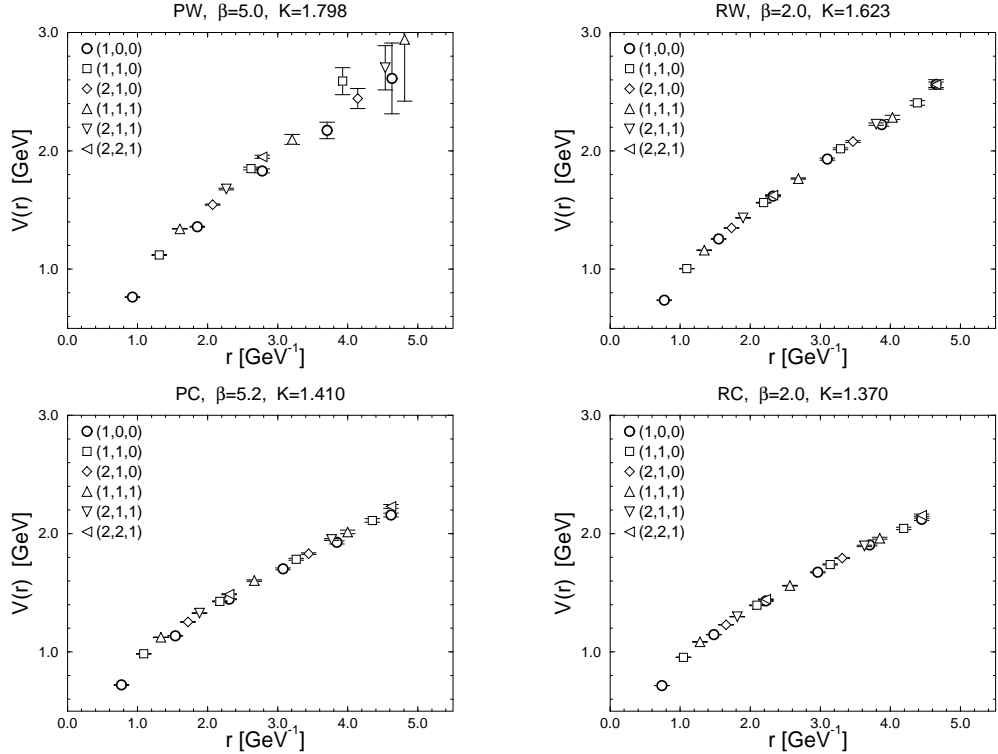

\vspace{-1mm}
\centerline{
\epsfxsize=6.8cm \epsfbox{Figs/VvsR.PW.b50K1798.mrhophys}
\hspace{3mm}
\epsfxsize=6.8cm \epsfbox{Figs/VvsR.RW.b20K1623.mrhophys}
}
\vspace{-1mm}
\centerline{
\epsfxsize=6.8cm \epsfbox{Figs/VvsR.PC.b52K1410.mrhophys}
\hspace{3mm}
\epsfxsize=6.8cm \epsfbox{Figs/VvsR.RC.b20K1370.mrhophys}
}
\vspace{-12mm}
\caption{
Static quark potential for the \PW, \RW, \PC, and \RC\ actions 
at $m_\pi/m_\rho \approx 0.8$.  Scales are normalized by 
the lattice spacing determined from $m_\rho$ in the 
chiral limit.
}
\label{fig:pot}
\vspace{-4mm}
\end{figure*}

\section{Light hadron masses}
\label{sect:mass}
\vspace{-1mm}

Our main results for the effect of improved actions on hadron masses 
are displayed in Fig.~\ref{fig:EPN} in which the ratio $m_N/m_\rho$ 
and $m_\Delta/m_\rho$ are plotted 
as a function of $(m_\pi/m_\rho)^2$ for the four
action combinations.  
The solid curves represent the well-known phenomenological mass 
formula \cite{ONO}.  
The inverse lattice spacing estimated from the $\rho$ meson mass in the 
chiral limit is in the 
range $a^{-1}\approx 1.1$-1.3GeV (see Table~\ref{tab:results}).  

For the standard action combination \PW, the ratios are well above the 
phenomenological curve as may be expected at such a large lattice spacing.
When we improve the gauge action (the \RW\ case), 
the data points come closer to the curve.
By far the most conspicuous change, however,  is observed
when we introduce the clover term to the quark action. 
For both the \PC\ and \RC\ cases, the data points drop significantly 
and lie on top of the phenomenological curve within errors.  
The same trend is seen both for $m_N/m_\rho$ and $m_\Delta/m_\rho$.

\begin{table} [t]
\caption{Results of $a^{-1}$ and $J$ determined by 
a linear fit of $m_\rho a$ in terms of $(m_\pi a)^2$ 
using data at $(m_\pi a)^2 \approx 0.3$-1.1
(see discussions in the text). 
Errors are statistical only. 
}
\label{tab:results}
\vspace{-0mm}
\begin{center}  
\begin{tabular}{clll} 
\hline
action & $\beta$ & $a^{-1}$[GeV] & $J$ \\
\hline
\PW\ & 4.8  &  1.00(1) & 0.35(1) \\ 
     & 5.0  &  1.08(2) & 0.35(1) \\ 
\RW\ & 1.9  &  1.15(1) & 0.34(1) \\
     & 2.0  &  1.29(1) & 0.34(1) \\
\PC\ & 5.2  &  1.30(3) & 0.42(1) \\
     & 5.25 &  1.48(6) & 0.41(2) \\
\RC\ & 2.0  &  1.35(1) & 0.44(1) \\
\hline
\end{tabular}
\end{center}
\vspace{-5mm}
\end{table}

\begin{figure*}[t]
\vspace{-1mm}
\centerline{
\epsfxsize=6.8cm \epsfbox{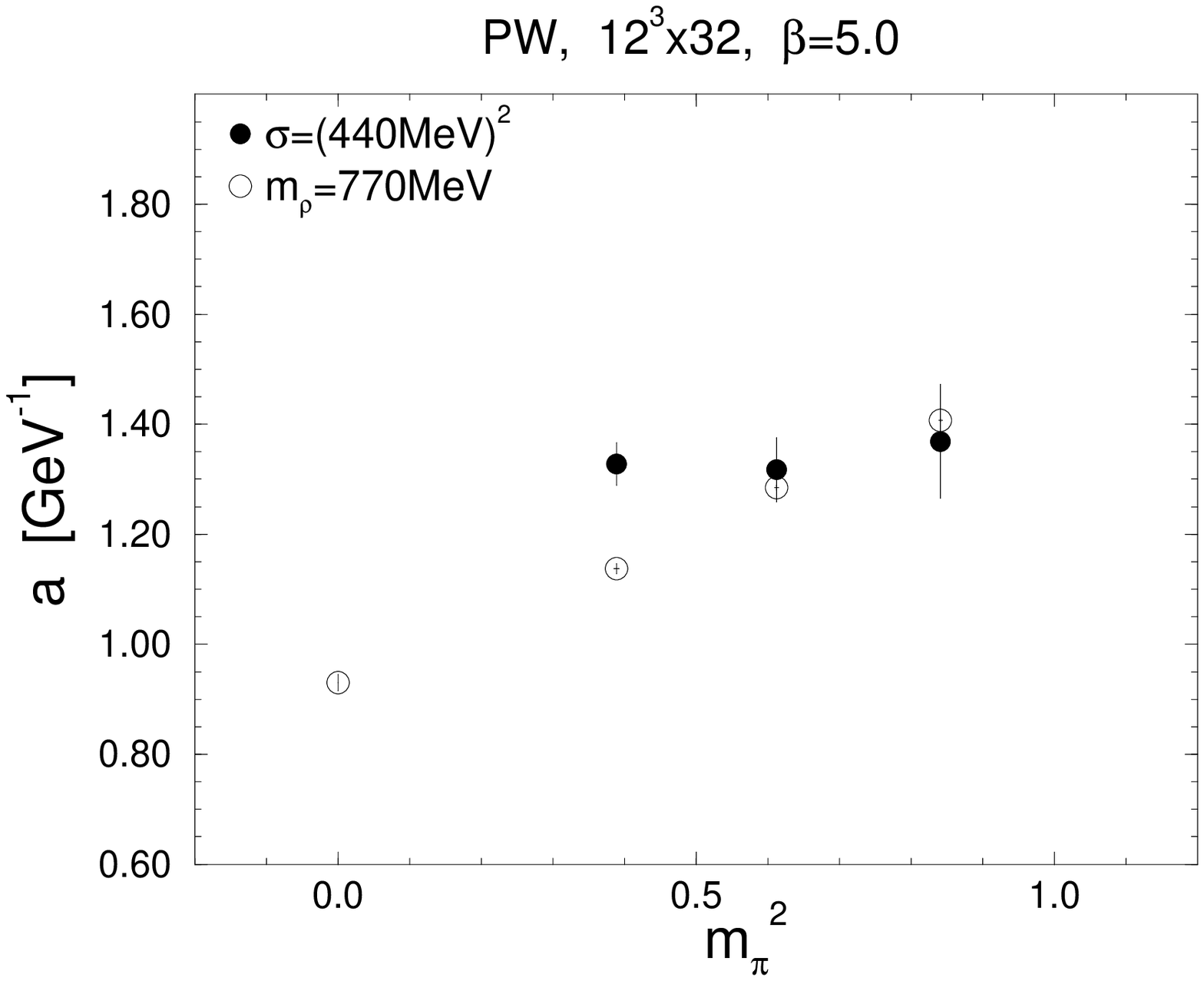}
\hspace{3mm}
\epsfxsize=6.8cm \epsfbox{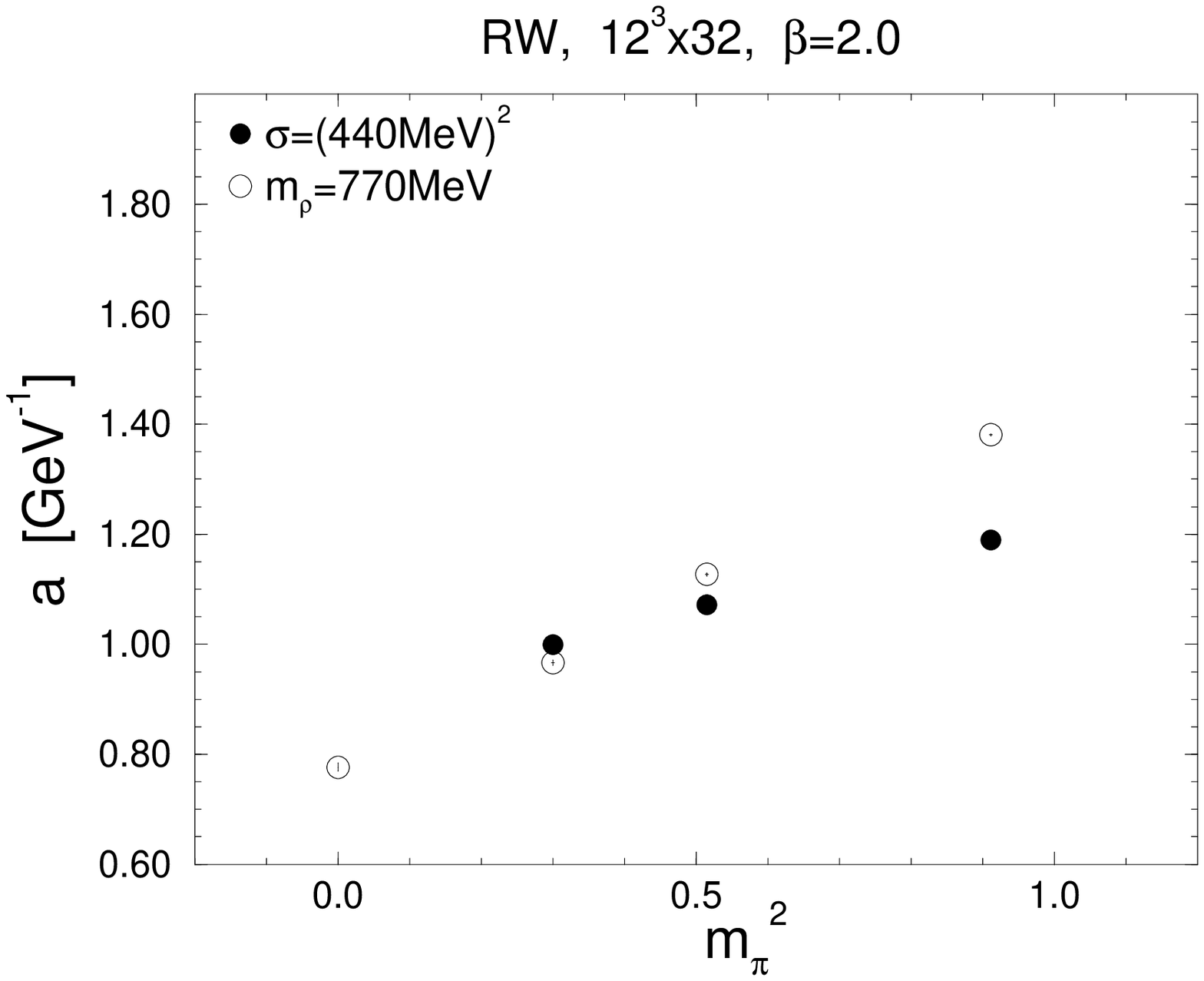}
}
\vspace{-1mm}
\centerline{
\epsfxsize=6.8cm \epsfbox{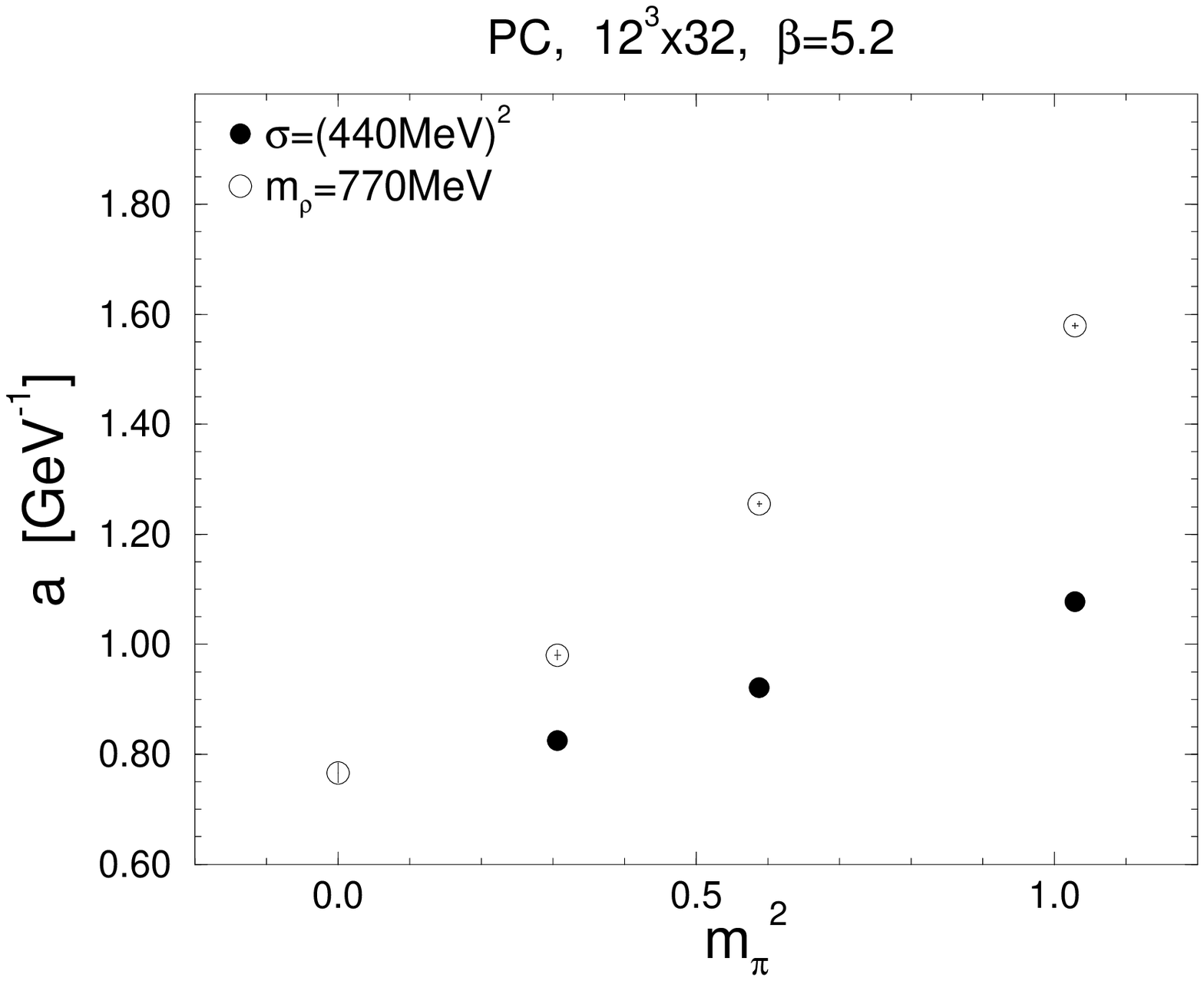}
\hspace{3mm}
\epsfxsize=6.8cm \epsfbox{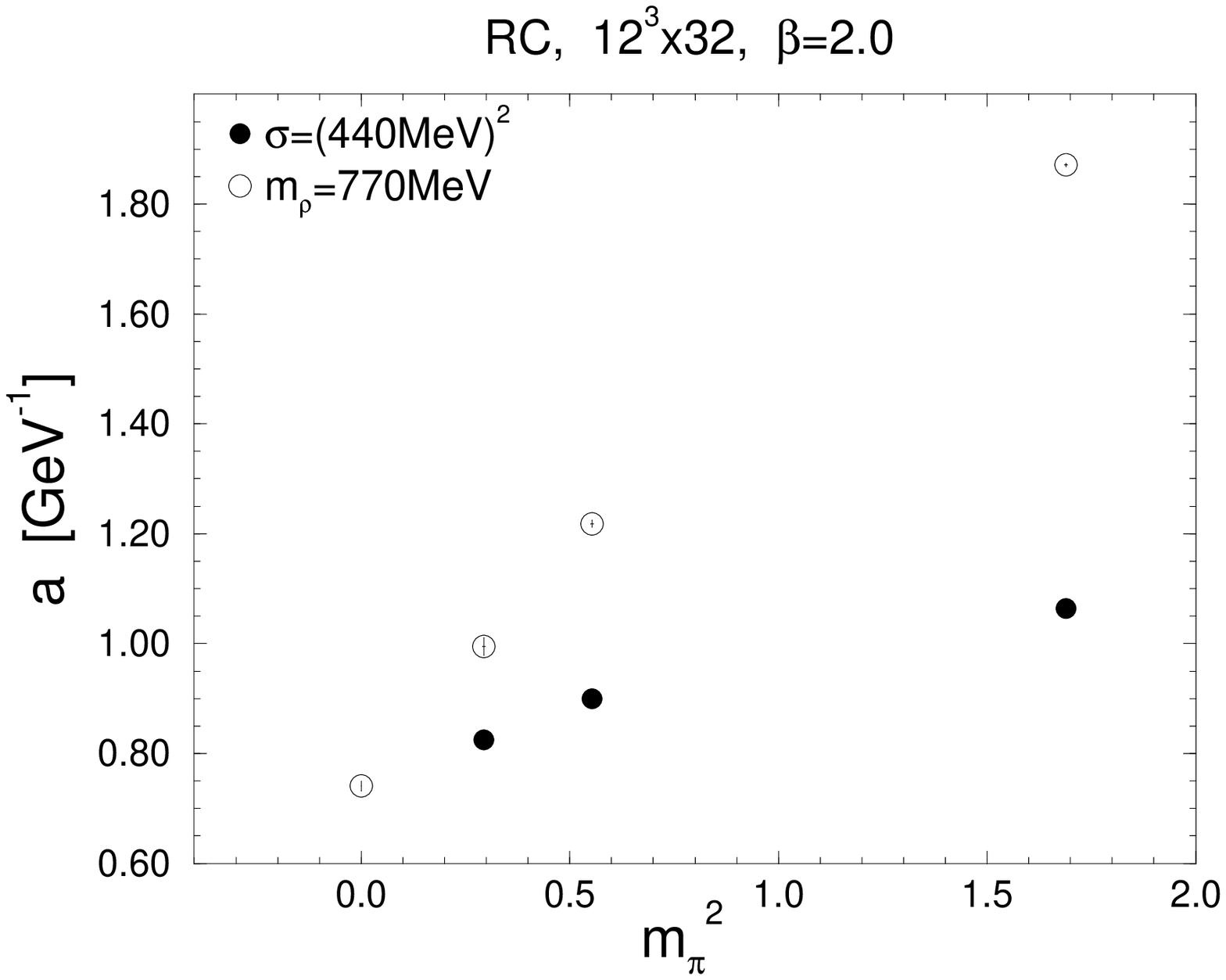}
}
\vspace{-12mm}
\caption{
Lattice spacing $a$ in GeV$^{-1}$ as a function of $(m_\pi a)^2$, 
for \PW, \RW, \PC, and \RC\ actions at 
$m_\pi/m_\rho \approx 0.8$. 
Note the difference in the scale of horizontal axis for {\sf R-C}.
}
\label{fig:a}
\vspace{-4mm}
\end{figure*}

A similar effect has been observed in quenched QCD
by the UKQCD Collaboration in simulations with the \PC\ combination 
at $\beta=5.7$-6.2 ($a^{-1}\approx 1.1$-2.5GeV) \cite{UKQCDKphi}, 
and also in a study with improved gluon actions and the clover and D234 
quark actions at coarse lattice spacings of $a^{-1}\approx 0.5$-0.7GeV
\cite{bock}.  It is therefore more natural to associate the origin of
the effect to valence quarks rather than to dynamical sea quarks in full QCD.
Nonetheless, changing the gluon and quark actions one at a time, we have 
been able to see clearly a decisive role played by the clover term in 
improving the 
spectrum in a quantitative detail in the context of full QCD.  In this 
regard, improving the gluon action has much less effect.

In Table~\ref{tab:results} we compile our results for the
$J$ parameter, $J=m_V \, dm_V / dm_{PS}^2$ 
at $m_V/m_{PS}=1.8$ \cite{UKQCDJ}.
Here again the clover term brings the values 
into better agreement with the experimental value of 0.48(2).

Another interesting feature in our hadron mass data is that 
they exhibit a negative curvature in terms of $1/K$ toward the chiral limit. 
This is in contrast to quenched QCD where hadron masses (mass squared for 
pseudo scalar mesons) are generally well described by a linear function 
of $1/K$ up to quite heavy quark.  The curvature is reduced
if hadron masses are plotted against $m_\pi^2$, but still remains at a 
significant level, especially for the \RC\ combination.  This is possibly 
a full QCD effect due to sea quarks which increasingly ordered gauge 
configurations toward the chiral limit.  
For results with clover quark actions the 
trend may be enhanced by the dependence of $c_{SW}$ on $K$
due to tadpole improvement.

The curvature causes a practical difficulty in the chiral extrapolation of 
hadron masses since our runs have so far been made with at most three 
values of $K$. For the estimates of $a^{-1}$ in Table~\ref{tab:results} 
we employed a linear fit of hadron masses in the measured values of 
$m_\pi^2$, excluding for the \RC\ case the point of heaviest quark mass.

In this connection we note that a comparison such as in 
Fig.~\ref{fig:EPN} and for 
$J$ in Table~\ref{tab:results} should be made
at the same lattice spacing in physical units.  
As we see in Table~\ref{tab:results} our estimate of $a^{-1}$ shows a 
spread of some 20-30\% depending on the actions.  Additional runs are 
being conducted in an effort to match the lattice spacing more precisely.

Another point to note is that an agreement with the  phenomenological 
formula of Ref.~\cite{ONO} examined in Fig.~\ref{fig:EPN} is not an 
improvement criterion that follows 
from theoretical principles, albeit a reasonable one in view of the  
success of the formula for describing the experimental spectrum.  
Stability of results toward smaller lattice spacing has to be 
checked, which we hope to pursue in future simulations.

\section{Static quark potential}
\label{sect:potential}
\vspace{-1mm}

We plot typical results for the static quark potential in Fig.~\ref{fig:pot}
for which $m_\pi/m_\rho \approx 0.8$. 
Conversion to physical units is made with the lattice scale 
given in Table~\ref{tab:results}.  
Different symbols correspond to potential data measured in different spatial 
directions along the vector given in the figure.

For the \PW\ action rotational symmetry is badly violated.  
Improving the quark action (\PC) we observe that the potential exhibits a 
much better rotational symmetry. 
At present how much of the improvement is due to the change of the quark 
action is not clear since the lattice spacing for the \PC\ 
case ($a^{-1}\approx 1.3$GeV) is about 20\% smaller than for the case of 
\PW\ ($a^{-1}\approx 1.1$GeV), from which one expects a 40\% reduction in the 
violation of rotational symmetry.  
The best improvement is achieved with the \RW\ and 
\RC\ actions.  It is known that the
gauge action \R\ significantly improves rotational 
symmetry already for the quenched case \cite{IwasakiPot97}. 
The improvement naturally carries over to the present case of full QCD.

With our present statistics the potential can be measured with small 
errors up to a distance of $r \approx 6a \approx 1$fm.  In this 
distance range we do not observe flattening of the potential due to 
pair creation and annihilation effects.  In fact the potential can be 
well described by a linear plus Coulomb form $\sigma r +\alpha/r$.  
We extract the string tension $\sigma$ by fitting potential data to 
this form.
We then use the phenomenological value 
$\sigma=(440{\rm MeV})^2$ to convert results to an estimate of $a$ 
for each value of $K$.

In Fig.~\ref{fig:a} we compare the value of $a$ obtained in this 
way (filled circles) with that from $m_\rho=770$MeV, 
also calculated for each value of $K$ (open circles). 
For the latter quantity, results obtained after an extrapolation 
of $m_\rho a$ to the chiral limit are also shown. 
This extrapolation is linear for $a$ when we take 
$(m_\pi a)^2$ as the horizontal axis, as adopted in Fig.~\ref{fig:a}.
We observe that the two estimates converge to a consistent value 
in the chiral limit for the \PC\ and \RC\ combinations, while 
an apparent deviation of order 40\% and 20\% are indicated 
for the cases of \PW\ and \RW\, respectively.  

Of course we expect the discrepancy observed for the last two cases 
with the Wilson quark action to disappear in the continuum limit.  
The agreement found for the \PC\ and \RC\ cases show that the clover 
term helps improve the consistency of the two determinations of the 
physical scale of lattice spacing already at $a^{-1}\approx 1.3$GeV.

The static quark potential at short distances is directly 
relevant to the spectrum 
of heavy quark systems.  A recent NRQCD calculation carried out 
on gauge configurations generated with the Kogut-Susskind action for sea 
quark and the plaquette gluon action at $\beta=5.6$ shows a significant 
deviation between the scale determined from the heavy and light hadron 
spectrum\cite{Sloan97}.  It would be interesting to check 
for our \PC\ and \RC\ cases if the lattice scale determined from heavy 
quark systems yield results consistent with those from $\sigma$ 
and $m_\rho$ .

\section{Conclusions}
\vspace{-1mm}

Our comparative study has shown that the clover term of 
Sheikholeslami and Wohlert
drastically improves light hadron spectrum already 
at $a^{-1} \approx 1.1$-1.3GeV.  While improving the gauge action 
has much less effect in this regard, we expect good rotational 
symmetry of static quark potential, which is achieved with the action \R,  
to be important for heavy quark systems.

Our results lead us to believe that a significant step forward towards 
a realistic simulation of full lattice QCD, encompassing 
both heavy and light hadrons,  can be 
achieved with the current generation of dedicated parallel computers with 
the application of improved actions.  We are particularly encouraged to pursue 
the use of the combination \RC\ toward this goal.

This work is supported in part by the Grant-in-Aid
of Ministry of Education, Science and Culture
(Nos.\ 08NP0101, 08640349, 08640350, 08640404, 08740189, 
and 08740221).  Three of us (GB, RB, and TK) are supported by 
the Japan Society for the Promotion of Science.


\begin{thebibliography}{9}

\bibitem{NRQCD97}
C.T.H.\ Davies {\it et al.}, 
Phys.\ Lett.\ B345 (1995) 42;  
C.T.H.\ Davies, these proceedings.

\bibitem{UKQCDJ}
P.\ Lacock and C.\ Michael,
Phys.\ Rev.\ D52 (1995) 5213.

\bibitem{LosAlamos96}
T.\ Bhattacharya {\it et al.}, 
Phys.\ Rev.\ D53 (1996) 6486.

\bibitem{UKQCDKphi}
R.\ Kenway for UKQCD Collaboration, 
Nucl.\ Phys.\ B (Proc.\ Suppl.) 53 (1997) 206.

\bibitem{YoshieCCP}
CP-PACS Collaboration (presented by T.\ Yoshi\'e), these proceedings.

\bibitem{UKQCDkenway} See, however, R. Kenway, these proceedings.

\bibitem{cppacs} 
Y.\ Iwasaki, these proceedings.

\bibitem{FukugitaAoki}
M.\ Fukugita {\it et al.}, 
Phys.\ Rev.\ D47 (1993) 4739;
S.\ Aoki {\it et al.}, {\it ibid.} D50 (1994) 486.

\bibitem{MILC_fullKS}
S. Gottlieb, these proceedings;
Nucl.\ Phys.\ B (Proc.\ Suppl.) 53 (1997) 155.

\bibitem{GF11}
F.\ Butler{\it et al.}, 
Nucl.\ Phys.\ B421 (1994) 217.

\bibitem{Sloan97}
J.\ Sloan, those proceedings.

\bibitem{cornell} M. Alford {\it et al.}, Phys. Lett. B361 (1995) 87;
G. P. Lepage, these proceedings.

\bibitem{scri}S. Collins {\it et al.}, 
Nucl. Phys. B(Proc. Suppl.) 47 (1996) 378;
{\it ibid.} 53 (1997) 877.

\bibitem{bock} W. Bock, Nucl. Phys. B(Proc. Suppl.) 53 (1997) 870.

\bibitem{ImQQcdMILC}
C.\ Bernard {\it et al.}, Nucl.\ Phys.\ B (Proc.\ Suppl.) 53 (1997) 212.

\bibitem{ImFQcdSCRI}
S.\ Collins {\it et al.}, 
Nucl.\ Phys.\ B (Proc.\ Suppl.) 47 (1996) 378;
{\it ibid.} 53 (1997) 880.

\bibitem{Iwasaki83}
Y.\ Iwasaki, Nucl.\ Phys.\ B258 (1985) 141;
Univ.\ of Tsukuba report UTHEP-118 (1983), unpublished.

\bibitem{IwasakiPot97}
Y.\ Iwasaki {\it et al.}, 
Nucl.\ Phys.\ B (Proc.\ Suppl.) 53 (1996) 429;
Phys.\ Rev.\ D56 (1997) 151. 

\bibitem{TIPot}
M.\ Alford, W.\ Dimm, and G.P.\ Lepage,
Nucl.\ Phys.\ (Proc.\ Suppl.) 42 (1995) 787.

\bibitem{FPPot}
T.\ DeGrand {\it et al.}, 
Nucl.\ Phys.\ B454 (1995) 615.

\bibitem{clover}
B.\ Sheikholeslami and R.\ Wohlert, Nucl.\ Phys.\ B259 (1985) 572.

\bibitem{meanfield}
G.P.\ Lepage and P.B.\ Mackenzie, Phys.\ Rev.\ D48 (1993) 2250;
P.B.\ Mackenzie, Nucl.\ Phys.\ B (Proc.\ Suppl.) 30 (1993) 35.

\bibitem{StdST}
G.S.\ Bali and K.\ Schilling, Phys.\ Rev.\ D46 (1992) 2636.

\bibitem{ONO}
S.\ Ono, Phys.\ Rev.\ D17 (1978) 888. 


\end{thebibliography}
\end{document}